\documentclass[12pt]{article}
\usepackage{amsfonts}
\usepackage{latexsym}
\usepackage{dsfont}
\usepackage{amsmath}
\usepackage{amssymb}
\usepackage{amssymb}
\usepackage{amsthm}
\usepackage[nosort]{cite}
\usepackage{setspace}
\usepackage{xcolor}
\usepackage{graphicx}
\usepackage{accents}
\usepackage[font=small,labelfont=bf]{caption}
\usepackage[utf8]{inputenc}
\usepackage{geometry}
\usepackage[toc,page]{appendix}

\usepackage[linktocpage=true]{hyperref}
\hypersetup{
colorlinks=false,
citecolor= blue,
linkcolor= blue,
urlcolor= blue,
breaklinks=true
}
\usepackage{cleveref}
\crefformat{section}{\S#2#1#3} 
\crefformat{subsection}{\S#2#1#3}
\crefformat{subsubsection}{\S#2#1#3}

\def\appendix#1{\addtocounter{section}{1}\setcounter{equation}{0}
\renewcommand{\thesection}{\Alph{section}}
\section*{Appendix \thesection\protect\indent \parbox[t]{11.15cm}{#1}}
\addcontentsline{toc}{section}{Appendix \thesection\ \ \ #1}}

\allowdisplaybreaks
\numberwithin{equation}{section}

\def\dd{\text{d}}
\def\lp{\lambda_+}
\def\lm{\lambda_-}

\begin{document}


\begin{titlepage}
\begin{center}
\vspace*{-1.0cm}
\hfill {\footnotesize HU-EP-22/09}

\vspace{2.0cm}

{\LARGE  {\fontfamily{lmodern}\selectfont \bf Coset space actions for nonrelativistic strings}} \\[.2cm]

\vskip 2cm
\textsc{Andrea Fontanella\footnote{\href{mailto:a.fontanella.physics@gmail.com}{\texttt{a.fontanella.physics@gmail.com}}} \footnotesize
and \normalsize Stijn J. van Tongeren\footnote{\href{mailto:svantongeren@physik.hu-berlin.de}{\texttt{svantongeren@physik.hu-berlin.de}}}\footnotesize }\\
\vskip 1.2cm

\begin{small}
\textit{Institut f\"ur Physik, Humboldt-Universit\"at zu Berlin, \\
IRIS Geb\"aude, Zum Gro{\ss}en Windkanal 2, 12489 Berlin, Germany}

\end{small}

\end{center}

\vskip 2 cm
\begin{abstract}
\vskip1cm
We formulate the stringy nonrelativistic limits of the flat space and AdS$_5\times$S$^5$ string as coset models, based on the string Bargmann and extended string Newton-Hooke algebras respectively. Our construction mimics the typical relativistic one, but differs in several interesting ways. Using our coset formulation we give a Lax representation of the equations of motion of both models.
\end{abstract}

\end{titlepage}

\tableofcontents
\vspace{5mm}
\hrule


 \setcounter{section}{0}
 \setcounter{footnote}{0}

 \section*{Introduction}

Non-relativistic (NR) string theory has attracted considerable attention in recent years, both as an alternate angle of approach to quantum gravity \cite{Hansen:2019vqf}, as well as an interesting area for holography \cite{Bagchi:2009my,Bagchi:2010zz} and non-relativistic limits of conventional AdS/CFT \cite{Harmark:2017rpg, Harmark:2018cdl, Harmark:2020vll}. The study of NR strings was initiated in \cite{Gomis:2000bd} via a particular infinite speed of light limit\footnote{This is not a strict limit, since divergences appear which require a nontrivial rewriting that introduces Lagrange multiplier fields in the action.} of the relativistic flat space string, see also \cite{Danielsson:2000gi}, then similarly investigated for the AdS$_5\times$S$^5$ string \cite{Gomis:2005pg}, and has recently been extended to general target space geometries \cite{Bergshoeff:2018yvt,Bidussi:2021ujm}.
Because NR string theory is naturally defined on a non-Lorentzian background it is an appealing arena to test non-AdS holography \cite{Sakaguchi:2007ba}. Moreover, NR string theory appears to be the natural UV completion of NR theories of gravity, e.g. \cite{VandenBleeken:2017rij, Hansen:2020pqs, Bergshoeff:2021bmc, Bergshoeff:2021tfn}.
The study of NR string theory includes Weyl anomalies \cite{Gomis:2019zyu, Gallegos:2019icg}, action symmetries \cite{Bergshoeff:2018yvt, Harmark:2019upf, Bergshoeff:2019pij}, connection to double field theory \cite{Ko:2015rha, Blair:2019qwi, Blair:2020gng, Blair:2020ops}, classical solutions \cite{Gomis:2004ht, Roychowdhury:2019olt,Fontanella:2021btt}, Hamiltonian formalism \cite{Kluson:2017abm,Kluson:2018egd,Kluson:2018grx}, and open strings \cite{Gomis:2020fui,Gomis:2020izd}, see e.g. \cite{Oling:2022fft} for a recent review.

In this paper we give a coset model formulation for the bosonic sectors of the stringy NR limit of the flat space \cite{Gomis:2000bd} and  AdS$_5\times$S$^5$ strings \cite{Gomis:2005pg}, which from here on out we will call the NR flat space and NR AdS string respectively. We also use this coset formulation to give a Lax representation of the corresponding equations of motion. Our formulation is based on the string Bargmann algebra and the direct sum of the extended string Newton-Hooke and Euclidean algebras, in the flat space and AdS cases, respectively. Related coset models were previously studied in \cite{Brugues:2004an,Brugues:2006yd}, however our formulation uses a different coset structure, and we moreover work with the Polyakov form of these strings to remain as close as possible to models such as the AdS$_5\times$S$^5$ string.\footnote{Similar coset structures were investigated for NR particle geometries in \cite{Grosvenor:2017dfs, Figueroa-OFarrill:2018ilb}.} Integrability of NR strings has previously been investigated in \cite{Kluson:2017ufb,Roychowdhury:2019vzh}, and for related sigma models in \cite{Fontanella:2020eje}, but not for the NR limit of the AdS$_5\times$S$^5$ string considered in \cite{Gomis:2005pg}.\footnote{The author of \cite{Roychowdhury:2019vzh} moreover appears to claim integrability for arbitrary (torsion-free) backgrounds, which we do not expect to be able to replicate.}

Until recently the NR string action for general target space geometry was formulated based on a gauging of the string Newton-Cartan (SNC) algebra \cite{Bergshoeff:2019pij}, using a particular torsion-free constraint.\footnote{There were also previous iterations built on the string Bargmann algebra, see e.g. \cite{Bergshoeff:2018yvt}.} Half a year ago this formulation was superceded by the so-called torsional string Newton-Cartan (TSNC) formulation \cite{Bidussi:2021ujm}, circumventing what may be viewed as downsides of the SNC formulation, such as field redundancy and associated St\"{u}ckelberg symmetry.\footnote{The possibility to avoid Stueckelberg type redundancies, as well as to take nonrelativistic limits without imposing torsion constraints mentioned below, were previously observed in NS-NS gravity in \cite{Bergshoeff:2021bmc}.}  As will come back shortly, the torsional string Newton-Cartan formulation uses a gauging procedure based on the so-called F string Galilei (FSG) algebra,\footnote{As the name implies, it imposes no constraints on torsion. The role of the torsion constraint was also investigated in \cite{Yan:2021lbe}.} which arises as a direct infinite speed of light limit of relativistic string symmetries. At the relativistic level, the string geometry is first formulated via a gauging procedure based on the so-called string Poincar\'e algebra, which does not impose a torsion constraint, and incorporates the B field. This string Poincar\'e algebra is an extension of the Poincar\'e algebra by a second independent set of generators transforming in the fundamental representation of the Lorentz algebra, with the same semi-direct product structure. These extra generators are associated to the B field of the string. The limiting procedure to arrive at the NR string involves mixing up the string Poincar\'{e} generators, and as a result there is no longer a semi-direct product structure involving the FSG generators associated to the TSNC two form.

Our goal is to formulate the NR flat space and AdS  strings in Polyakov form as coset sigma models, mirroring the construction of relativistic symmetric space sigma models such as the AdS$_5\times$S$^5$ string \cite{Metsaev:1998it}, see e.g. \cite{Arutyunov:2009ga} for a review. In the relativistic setting we can naturally embed torsion-free geometry in the string Poincar\'{e} picture mentioned above, since in essence we can simply drop the additional generators corresponding to the B field.\footnote{Put differently, restricting the algebra effectively sets the B field to zero, but the interpretation of the vielbein remains the same.} Hence we can use (torsion-free) Maurer-Cartan (MC) geometry as the natural language to describe a relativistic string with a symmetric target space. In contrast, in the NR setting, the torsional nature of the TSNC formulation appears to be at odds with MC type language. Moreover, at least for the  NR flat space string, there is a concrete obstacle. A coset formulation is based on the global symmetry algebra of a model, which for the NR flat space string was determined in \cite{Batlle:2016iel}.  However, it appears impossible to embed a sufficiently large subalgebra of the FSG in this global symmetry algebra.\footnote{E.g. assuming a standard embedding of the Galilean generators, the Galilean boost generators in FSG commute, while they do not in the global symmetry algebra of \cite{Batlle:2016iel}.} In other words, the TSNC analogue of relativistic local Lorentz symmetry has no clear relation to an $H$ we would pick in a coset formulation for the NR flat space string. This obscures the interpretation of the components of a MC type form in terms of the geometric data of the TSNC formulation.

Coming back to our actual coset sigma model construction, we first consider the NR flat space string. The global symmetry algebra of this string is actually infinite dimensional \cite{Batlle:2016iel}, leading us to consider a suitable finite dimensional subalgebra as the basis for our coset model instead. There are two natural candidates for such a subalgebra \cite{Bergshoeff:2019pij}: the string Bargmann and the SNC algebra. We find that while both choices are admissible, the string Bargmann choice is more economical, and appears naturally adapted to finding a Lax formulation of the equations of motion. Our coset model construction is different from the standard relativistic one however. In a standard $G/H$ coset sigma model, local $H$ invariance follows essentially from $H$ invariance of the bilinear form on the (relevant subspace of) the Lie algebra of $G$. In our case, this partly works as usual, but also partly relies on a cancellation between gauge transformations of the extra Lagrange multiplier fields that appear in the Polyakov formulation of the NR string, and gauge transformations of the usual MC action arising from a not fully $H$ invariant bilinear form. As it turns out, it proves practical to refer to the SNC formulation of the NR string to see this, if only for reasons of familiarity.

Due to our differently realized gauge symmetry, naively varying the action gives more nontrivial equations of motion than independent fields, but these are appropriately related by corresponding Noether identities. Moreover, based on our coset formulation and the form of the equations of motion, there is a simple argument that allows us to modify the standard Lax connection encoding the equations of motion of a relativistic symmetric space sigma model, and obtain a Lax formulation for the present case.\footnote{This establishes integrability in the so-called weak sense. To establish integrability in the strong sense -- sufficiently many conserved charges in involution -- requires a careful analysis in the Hamiltonian setting, whereby the expression for the Lax connection might need to be modified in particular by constraints, see e.g. \cite{Magro:2008dv,Vicedo:2009sn}. See \cite{Kluson:2018grx} for a hamiltonian analysis of the NR SNC string.}

Coming finally to the NR AdS string, we find that while its (infinite dimensional) global symmetry algebra has not been explicitly determined, the extended string Newton-Hooke algebra provides a suitable analogue of the string Bargmann algebra used to construct the flat space coset model, requiring only minimal changes to the derivation of the coset action. Here our Lax representation of the equations of motion still works, now on solution of the constraints enforced by the Lagrange multipliers.

\textbf{Structure of the paper}. In \cref{sec:NR_flat} we give a coset formulation of the NR string action in flat space and a Lax representation for its equations of motion. In \cref{sec:NR_AdS} we repeat the construction for NR strings in AdS. The paper ends with several appendices. In \cref{app:Conventions} we give our conventions and provide useful identities. In \cref{app:SNCreview} we review SNC strings, make contact with the coset construction and review their connection with relativistic strings. In \cref{app:coordinateandMCeom} we show the relation between the equations of motion in the coset and Polyakov languages. In \cref{app:Liealgebraexp} we derive the global symmetry algebra of AdS strings. In \cref{app:SNCcosetconstruction} we provide the coset construction of NR strings in flat space based on the SNC algebra.

\section{The NR flat space string as a coset model}\label{sec:NR_flat}

The action for the NR flat space string in Polyakov form is given by \cite{Gomis:2000bd}
\begin{equation}\label{eq:NRflat}
S =  - \frac{T}{2} \int \dd^2 \sigma \, \gamma^{\alpha\beta}\partial_{\alpha} x^{a} \partial_{\beta} x^{b} \delta_{ab} + \varepsilon^{\alpha\beta} (\lp e_{\alpha}{}^+ \partial_\beta x^+ + \lm e_{\alpha}{}^- \partial_\beta x^- ) \,,
\end{equation}
where $T$ is the string tension, $\sigma^{\alpha} = (\tau, \sigma)$, with $\alpha = 0, 1$, are the string world-sheet coordinates, $\gamma^{\alpha\beta} \equiv \sqrt{-h} h^{\alpha\beta}$ is the Weyl invariant combination of the inverse world-sheet metric $h^{\alpha\beta}$ and $h =$ det$(h_{\alpha\beta})$,and $e_{\alpha}{}^{\pm}$ are (the light-cone components of) the worldsheet zweibein, see appendix \ref{app:Conventions} for our conventions. The $x^{a}$, $a=2,\ldots,9$ are the transverse string embedding coordinates, and $x^\pm$ the longitudinal ones. $\lp$ and $\lm$ are non-dynamical scalar Lagrange multiplier fields.

Integrating out the Lagrange multiplier fields makes the worldsheet metric conformally flat, and in static gauge ($x^\pm = \tau \pm \sigma$) gives the corresponding Nambu-Goto action \cite{Gomis:2000bd}
\begin{equation}\label{eq:NRflatNG}
S_{NG}= S =  - \frac{T}{2} \int \dd^2 \sigma \, \eta^{\alpha\beta}\partial_{\alpha} x^{a} \partial_{\beta} x^{b} \delta_{ab}\,,
\end{equation}
describing eight free massless bosons.

To write down a $G/H$ coset model corresponding to this string, we need its global symmetry group $G$, an appropriate denominator $H$, and a suitable bilinear form on the Lie algebra $\mathfrak{g}$ of $G$. As our construction will necessarily differ from the typical relativistic symmetric space sigma model construction, let us briefly recall this for illustrative purposes.

\subsection{Relativistic coset models}

Let us consider a (pseudo-)Riemannian symmetric space $M$, i.e. Lie group theoretically a manifold that can be represented in the form $G/H$, where $G$ is the isometry group of $M$ and $H$ is the isotropy group of a point in $M$, with associated Lie algebras $\mathfrak{g}$ and $\mathfrak{h}$ respectively. The symmetric space structure means that $\mathfrak{g}$ has a $\mathbb{Z}_2$ graded structure, splitting as $\mathfrak{g}^{(0)} \oplus \mathfrak{g}^{(1)} = \mathfrak{h} \oplus \mathfrak{p}$ with respect to the corresponding automorphism. The string action on $M$ is built using a nondegenerate $\mathfrak{h}$-invariant grade-compatible symmetric bilinear form $\langle,\rangle$,\footnote{$G$ is typically taken (semi-)simple, where $\langle,\rangle$ is (can be taken proportional to) the Killing form, meaning it has full $\mathfrak{g}$ invariance.} and a Maurer-Cartan form $A = g^{-1} \dd g$, $g \in G$, as
\begin{equation}
\label{eq:relativisticcoset}
S = \int \dd^2 \sigma \, \gamma^{\alpha\beta} \langle A_\alpha, P A_\beta \rangle,
\end{equation}
where $P$ projects $\mathfrak{g}$ onto $\mathfrak{p}$. Due to the projector and the $\mathfrak{h}$-invariance of the bilinear form, the action has $H$ gauge invariance under right multiplication of $g$, and we describe a model on $G/H$.

We would now like to give a similar construction for the NR flat space string \eqref{eq:NRflat}, which starts with its global symmetry algebra.

\subsection{Symmetry algebra}
\label{sec:symmetry_flat}

The global symmetry algebra of the flat space NR string was determined in \cite{Batlle:2016iel}, and shown to be a particular infinite dimensional extension of the string Galilei algebra, see also \cite{Bergshoeff:2019pij,Gomis:2019fdh}. Since it seems unnecessary and undesirable to attempt to use an infinite dimensional group in a coset model construction, we will try to base ourselves on a suitable finite dimensional subalgebra instead. Two such candidate algebras are known \cite{Bergshoeff:2019pij}, the string Bargmann and the SNC algebra. Both algebras can be used for a coset model construction, we focus here on the simpler string Bargmann case, giving details on the SNC case in appendix \ref{app:SNCcosetconstruction}. The string Bargmann algebra is spanned by a longitudinal boost $M$, longitudinal translations $H_A$, transverse rotations $J_{ab}$, transverse translations $P_a$, string-Galilei boosts $G_{Ab}$, and non-central extensions $Z_A$ and $Z$, with commutation relations
\begin{equation} \label{eq:Bargmann}
	\begin{aligned}
	{}[J_{ab}, J_{cd}] & = \delta_{bc} J_{ad} - \delta_{ac} J_{bd} + \delta_{ad} J_{bc} - \delta_{bd} J_{ac} \,, \qquad &  [M,G_{Aa}]  & = -\varepsilon_A{}^B G_{Ba}\,, \\
    [J_{ab},P_{c}] & =  \delta_{bc} P_{a} - \delta_{ac} P_{b}\,, & [M,H_A] & = -\varepsilon_A{}^B H_B\,, \\
    [J_{ab},G_{Ac}] & = \delta_{bc} G_{Aa} - \delta_{ac} G_{Ab} \,,& [M,Z_A] & = -\varepsilon_A{}^B Z_B\,, \\
	[G_{Aa}, G_{Bb}] & = \delta_{ab} \varepsilon_{AB} Z\,, & [Z,H_A]  & = -\varepsilon_A{}^B Z_B\,,\\
	[G_{Aa}, P_{b}] & = \delta_{ab} Z_A\,,&&\\
	[G_{Aa},H_B]  & = -\eta^{}_{AB} P_{a}\,. &&
	\end{aligned}
\end{equation}
We take $G$ to be the corresponding string Bargmann group.

\subsection{Gauge group, bilinear form}

Next we need to determine our coset denominator $H$, with Lie algebra $\mathfrak{h}$. Here our construction starts to differ from the relativistic case.

We want to encode our dynamical degrees of freedom in a MC form. However, we have extra Lagrange multiplier fields that do not naturally fit this language, and require modifying the typical construction. Fortunately, the close link between the string-Bargmann algebra and the SNC formulation of the nonrelativistic string, gives us a clear path to construct our model. First there is the question how to choose $H$. By analogy to relativistic strings we take $H$ to be generated by everything except the $H_A$ and $P_a$, further supported by the association between these generators and the longitudinal and transverse vielbeine in the SNC picture, see appendix \ref{app:SNCreview} for a review of the SNC formulation.\footnote{Our $G/H$ hence does not have the structure of a (Riemannian) symmetric space -- $[\mathfrak{h},\mathfrak{p}] \not\subset \mathfrak{p}$.} Second, in the SNC formulation \cite{Bergshoeff:2019pij}, the Lagrange multiplier fields transform under $Z_A$ transformations, meaning that we should not expect $Z_A$ transformations of our dynamical (MC) fields to leave the action invariant separately. Consequently, we should not insist on full $\mathfrak{h}$ invariance for our bilinear form.

Concretely we take $\mathfrak{h} = \text{span} \{M, J_{ab}, G_{Aa}, Z_A, Z \}$, and look for a bilinear invariant under the adjoint action of $\tilde{\mathfrak{h}} \equiv \mathfrak{h}\setminus \{Z_A\}$, i.e.
\begin{equation}
\langle [T_a, t_i], T_b \rangle + \langle T_a ,[T_b, t_i] \rangle = 0 \ , \qquad
\forall \ T_a, T_b \in \mathfrak{g} \ , \quad
\forall\ t_i \in  \tilde{\mathfrak{h}}\ .
\end{equation}
The corresponding symmetric bilinear form is\footnote{The antisymmetrization on $\langle J_{ab}, J_{cd} \rangle$ should be read as antisymmetrization in $a$ and $b$, and $c$ and $d$, respectively. Similar notation is used elsewhere in the paper.}
\begin{subequations}\label{inner_product}
	\begin{align}
	\langle P_a, P_b \rangle &= \omega_1  \delta_{ab} \ , &
	\langle H_{\pm}, Z_{\mp} \rangle &=  -\omega_1/2 \ , \\
	\langle H_+, H_- \rangle &= \omega_2 \ , &
	\langle M, M \rangle &= \omega_3 \ , \\
	\langle J_{ab}, J_{cd} \rangle &= \omega_4 \delta_{[a[c} \delta_{b]d]}  \ ,
	\end{align}
\end{subequations}
where the $\omega_i$ are arbitrary coefficients. It is degenerate on the full string Bargmann algebra -- $\langle Z, \cdot\rangle =0$ for example -- but nondegenerate on the span of $\{ H_A, P_a, Z_A \}$ (for nonzero $\omega_1$), which is sufficient.

Out bilinear form has further bonus invariance conditions
\begin{equation}
\label{bonus}
	\begin{aligned}
	\langle [H_+, x], H_- \rangle + \langle H_+,[H_-, x] \rangle &= 0 \ , \\
	\langle [Z_+, x], Z_- \rangle + \langle Z_+,[Z_-, x] \rangle &= 0 \ , \\
	\langle [P_a, x], P_b \rangle + \langle P_a,[P_b, x] \rangle &= 0 \ ,
	\end{aligned}
\end{equation}
for $x \in \{Z_{\pm}\}$.

\subsection{Action}
\label{sec:NRflatspaceaction}

To give an action for our coset model, we construct a MC form $A = g^{-1} \dd g$ with $g$ an element of the string Bargmann group, with algebra components
\begin{equation}
A = A^{H_A} H_A + A^{P_a} P_a + A^{Z_A} Z_A + A^{M} M + \frac{1}{2} A^{J_{ab}} J_{ab} + A^{G_{Aa}} G_{Aa} + A^{Z} Z \ ,
\end{equation}
and we collect the Lagrange multiplier fields in a separate current
\begin{equation}
\Lambda_{\alpha} = \Lambda_{\alpha}^{Z_A}\, Z_A  \equiv \lm e_{\alpha}{}^-  \, Z_+ + \lp e_{\alpha}{}^+  \, Z_- \ .
\end{equation}
To reproduce the flat space action we need to keep specific components of our MC form. Also here we could appeal to the SNC formulation to arrive at an appropriate choice. This choice turns out to be compatible with a grading that can be put on the string Bargmann algebra \eqref{eq:Bargmann}.\footnote{In contrast to the relativistic symmetric space case where the coset structure, grading and projector, and the required invariance, are all compatible, here the coset denominator is not exactly aligned with the remaining structure.} Namely, this algebra can be split as
\begin{equation}
\label{eq:grading}
\mathfrak{g} = \mathfrak{g}^{(0)} \oplus \mathfrak{g}^{(1)} \ , \qquad
\mathfrak{g}^{(0)} = \text{span} \{ M, J_{ab}, G_{Aa}, Z \}\ , \qquad
\mathfrak{g}^{(1)} = \text{span} \{ H_A, P_a, Z_A \}  \ ,
\end{equation}
where $\mathfrak{g}^{(0)}$ and $\mathfrak{g}^{(1)}$ are the eigenspaces of a $\mathbb{Z}_2$ automorphism $\Omega : \mathfrak{g} \rightarrow \mathfrak{g}$, corresponding to eigenvalues $1$ and $-1$ respectively. This grading is compatible with our bilinear form: defining $P$ to be the projector from $\mathfrak{g}$ down to $\mathfrak{g}^{(1)}$, we have $\langle x, P  y \rangle = \langle P x, P y \rangle$, for all $x,y\in \mathfrak{g}$. Note that $\mathfrak{g}^{(0)}$ is nothing but the $\tilde{\mathfrak{h}}$ that our bilinear form is invariant under. As mentioned above, our bilinear form is nondegenerate on $\mathfrak{g}^{(1)}$.

With this projector, fixing our bilinear form by taking $\omega_1 = 1$ and $\omega_2 = 0$, we can write the action for the NR flat space string as\footnote{See appendix \ref{app:SNCreview} for the link between this form and the SNC formulation of the NR string.}
\begin{equation}
\label{eq:NR_coset_action}
S^{\mathcal{G}/\mathcal{H}} = - \frac{T}{2} \int \dd^2 \sigma \,
\left( \gamma^{\alpha\beta}  \langle A_{\alpha}, P A_{\beta} \rangle
+ 2 \varepsilon^{\alpha\beta} \langle A_{\alpha}, P \Lambda_{\beta} \rangle \right) \ .
\end{equation}

Part of the $H$ gauge invariance of this action follows from the typical considerations for gauge invariance of the relativistic symmetric space model. Namely the action is manifestly invariant under right multiplication of $g \rightarrow g h$ combined with $\Lambda \rightarrow h^{-1} \Lambda h$ by elements $h$ of the group generated by $\tilde{\mathfrak{h}} = \mathfrak{g}^{(0)}$. This follows from the grading of the algebra, the projector, and grade compatibility and adjoint invariance of the bilinear form.

To have full $H$ gauge invariance, we also need invariance under transformations associated to $Z_A$. This gauge invariance is well known in the SNC formulation of the NR string, see appendix \ref{app:SNCreview} for details. In our language, the action is invariant under the transformation\footnote{While the transformation of $\Lambda$ involves algebra elements, they are not required to be infinitesimal -- no higher order terms are required for invariance.}
\begin{equation}
\begin{aligned}
\Lambda_{\alpha} & \rightarrow \Lambda_{\alpha} + 2 e_{\alpha}{}^+ e^\beta{}_+ D_\beta (\sigma^- Z_-) - 2 e_{\alpha}{}^- e^\beta{}_- D_\beta (\sigma^+ Z_+)\ ,\\
g & \rightarrow g e^{\sigma^A Z_A},
\end{aligned}
\end{equation}
where $D$ is the covariant derivative
\begin{equation}
\label{eq:covD}
D f = \dd f + [A,f]\ ,
\end{equation}
where the variations of the kinetic and Wess-Zumino term cancel up to a total derivative, via identities for the worldsheet zweibein given in appendix \ref{app:Conventions}, and using flatness of $A$.

Finally, the coset representative
\begin{equation}
g = e^{x^\pm H_\pm} e^{x^a P_a} \ ,
\end{equation}
and associated MC form
\begin{equation}
A = \dd x^\pm H_\pm + \dd x^a P_a \ ,
\end{equation}
explicitly give the coordinate action \eqref{eq:NRflat} above.

\subsection{Equations of motion}

We can compute the equations of motion associated to the action \eqref{eq:NR_coset_action} by considering an arbitrary variation of the group element $g$,  $\delta g = g \xi$, which induces $\delta A_{\alpha} = \partial_{\alpha} \xi+ [A_{\alpha}, \xi]$. We also vary the Lagrange multiplier fields, giving $\delta \Lambda_{\alpha} = e_{\alpha}{}^- \delta \lm \, Z_+ + e_{\alpha}{}^+ \delta \lp \, Z_-$. By splitting the variation of $\xi$ into its components, and using the invariance properties of our bilinear form, we get the following equations of motion
\begin{subequations} \label{Coset_eom}
\begin{align}
\label{P_eom}
 & \xi^P:  & \mathcal{E}^P \equiv &\ \partial_{\alpha}(\gamma^{\alpha\beta} A_{\beta}^{P_a}) + \gamma^{\alpha\beta}\left(\delta_{bc} A_{\alpha}^{J_{ab}} A_{\beta}^{P_c}  - \eta_{AB} A_{\alpha}^{G_{Aa}} A_{\beta}^{H_{B}} \right) = 0 \ , \\
 & \xi^{H_+}: &\mathcal{E}^{Z_-} \equiv &\ \partial_{\alpha}(\gamma^{\alpha\beta} A_{\beta}^{Z_-} + \varepsilon^{\alpha\beta} \Lambda_{\beta}^{Z_-})  - \varepsilon^{\alpha\beta} A_{\alpha}^M \Lambda_{\beta}^{Z_-} \hspace{3cm} \notag \\
 \label{H_eom}
 & \hspace{3cm}& &+ \gamma^{\alpha\beta} \left( \delta_{ab} A_{\alpha}^{G_{-a}} A_{\beta}^{P_b} - A_{\alpha}^M A_{\beta}^{Z_-} - A_{\alpha}^{Z} A_{\beta}^{H_-} \right) = 0 \ , \\
 & \xi^{H_+}: &\mathcal{E}^{Z_+} \equiv &\ \partial_{\alpha}(\gamma^{\alpha\beta} A_{\beta}^{Z_+} + \varepsilon^{\alpha\beta} \Lambda_{\beta}^{Z_+})  + \varepsilon^{\alpha\beta} A_{\alpha}^M \Lambda_{\beta}^{Z_+} \hspace{3cm} \notag \\
  \label{Hb_eom}
& \hspace{3cm}& &+ \gamma^{\alpha\beta} \left( \delta_{ab} A_{\alpha}^{G_{+a}} A_{\beta}^{P_b}+ A_{\alpha}^M A_{\beta}^{Z_+} + A_{\alpha}^{Z} A_{\beta}^{H_+} \right) = 0 \ , \\
 \label{Z_eom}
 &\xi^{Z_+}: & \mathcal{E}^{H_-} \equiv &\ \partial_{\alpha}(\gamma^{\alpha\beta} A_{\beta}^{H_-}) - \gamma^{\alpha\beta} A_{\alpha}^M A_{\beta}^{H_-}  = 0 \ , \\
 \label{Zb_eom}
 & \xi^{Z_-}: & \mathcal{E}^{H_+} \equiv &\ \partial_{\alpha}(\gamma^{\alpha\beta} A_{\beta}^{H_+}) + \gamma^{\alpha\beta}A_{\alpha}^M A_{\beta}^{H_+} = 0 \ , \\
 \label{M_eom}
 & \xi^{M}:  &\mathcal{E}^{M} \equiv &\ \varepsilon^{\alpha\beta} \left( \lp e_{\alpha}{}^+ A_{\beta}^{H_+} - \lm e_{\alpha}{}^- A_{\beta}^{H_-} \right) = 0 \ , \\
 \label{l+_eom}
 &\delta \lp:  &\mathcal{E}^{\lp} \equiv &\ \varepsilon^{\alpha\beta} e_{\alpha}{}^+ A_{\beta}^{H_+} = 0 \ , \\
 \label{l-_eom}
 &\delta \lm:  &\mathcal{E}^{\lm} \equiv &\ \varepsilon^{\alpha\beta}e_{\alpha}{}^- A_{\beta}^{H_-} = 0 \  .
\end{align}
\end{subequations}
We can represent these equations efficiently by introducing
\begin{equation}
J_{\alpha} \equiv A_{\alpha} - (\star \Lambda)_{\alpha} = A_{\alpha} + \gamma_{\alpha\beta}\varepsilon^{\beta\gamma}\Lambda_{\gamma} \ ,
\end{equation}
in terms of which equations (\ref{P_eom}), (\ref{H_eom}), (\ref{Hb_eom}), (\ref{Z_eom}) and (\ref{Zb_eom}) become the single algebra-valued equation
\begin{equation}
\label{eq:eom_compact}
\partial_{\alpha} (P \gamma^{\alpha\beta} J_\beta) + \gamma^{\alpha\beta} [J_{\alpha}, P J_\beta] = 0 \,.
\end{equation}

It may appear that at this stage that we have more equations of motion than we would expect for ten dynamical fields and two Lagrange multipliers. However, we also have gauge invariance at play. This leads to Noether identities between the equations of motion above, and appropriately reduces their number.\footnote{In the relativistic symmetric space sigma model, the gauge directions are nicely aligned with the structure of the action, and the equations of motion naturally split into physical and (identically vanishing) gauge directions, in line with the grading of the algebra. Here the structure is mixed up.} The relevant gauge transformations are those assocated to $Z_A$ and $M$.

First we consider the $Z_{\pm}$ gauge variation of the action, with gauge parameters $\sigma^{\pm}$. Gauge invariance of the action implies that
\begin{equation}
\delta_{\sigma^{\pm}} S^{\mathcal{G}/\mathcal{H}} = \frac{\delta S^{\mathcal{G}/\mathcal{H}}}{\delta A_{\mu}^{Z_{\pm}}} \delta_{\sigma^{\pm}} A_{\mu}^{Z_{\pm}} + \frac{\delta S^{\mathcal{G}/\mathcal{H}}}{\delta \lambda_{\mp}} \delta_{\sigma^{\pm}} \lambda_{\mp} \overset{!}{=} 0 \ ,
\end{equation}
which after an integration by parts gives the Noether identities
\begin{subequations} \label{Noether_Z_A}
   	\begin{align}
\mathcal{E}^{H_+} - 2 \mathcal{D}_{\alpha}( e^{\alpha}{}_+ \mathcal{E}^{\lp} ) &= 0 \ , \\
\mathcal{E}^{H_-} + 2 \mathcal{D}_{\alpha}( e^{\alpha}{}_- \mathcal{E}^{\lm} ) &= 0 \ ,
	\end{align}
\end{subequations}
where $\mathcal{D}$ is the component covariant derivative, $\mathcal{D}_{\alpha} f^A = \partial_\alpha f^A - \varepsilon_B{}^{A} A^M_\alpha f^B$.\footnote{\label{footnote:covD} This is $D$ of eqn. \eqref{eq:covD} with image and domain $\{H_A\}$. I.e., $\mathcal{D} f^{A} = (D f^C H_C)|_{H_A}$. Given the commutation relations, the same formula applies when we replace $H_A$ by $Z_A$, and can hence refer consistently to just the index $A$.} Next, the $M$ gauge variation of the action with parameter $\Sigma$
\begin{equation}
\delta_{\Sigma}  S^{\mathcal{G}/\mathcal{H}}= \frac{\delta S^{\mathcal{G}/\mathcal{H}}}{\delta A_{\mu}^{H_+}} \delta_{\Sigma} A_{\mu}^{H_+}
+\frac{\delta S^{\mathcal{G}/\mathcal{H}}}{\delta A_{\mu}^{H_-}} \delta_{\Sigma} A_{\mu}^{H_-}
+ \frac{\delta S^{\mathcal{G}/\mathcal{H}}}{\delta \lambda_+} \delta_{\Sigma} \lambda_+
+ \frac{\delta S^{\mathcal{G}/\mathcal{H}}}{\delta \lambda_-} \delta_{\Sigma} \lambda_- \overset{!}{=} 0 \ ,
\end{equation}
where only the Wess-Zumino term contributes nontrivially, is responsible for the nontrivial Noether identity
\begin{equation}\label{Noether_M}
\mathcal{E}^M + \lp \mathcal{E}^{\lambda_+} - \lm \mathcal{E}^{\lambda_-} = 0 \ .
\end{equation}

The Noether identities (\ref{Noether_Z_A}) and (\ref{Noether_M}) reduce the number of independent equations of motion. We can choose
\begin{equation}
\label{independent_set}
\mathcal{E}^P = 0 \ , \qquad
\mathcal{E}^{Z_{\pm}} = 0 \ , \qquad
\mathcal{E}^{\lambda_{\pm}} = 0 \ ,
\end{equation}
as our independent equations of motion. These are equivalent to the equations of motion derived directly from the component action \eqref{eq:NRflat}, see appendix \ref{app:coordinateandMCeom} for details.

\subsection{Lax representation}

The form of the equations of motion \eqref{eq:eom_compact} is suggestive. Compared to the equations of motion of the relativistic symmetric space sigma model \eqref{eq:relativisticcoset}
\begin{equation}
\mathcal{E} = \partial_{\alpha} (P \gamma^{\alpha\beta} A_\beta) + \gamma^{\alpha\beta} [A_{\alpha}, P A_\beta] =0
\end{equation}
the only difference with \eqref{eq:eom_compact} is that we have $A$ instead of $J$.\footnote{In the relativistic case the equations of motion typically immediately follow by using full $\mathfrak{g}$ invariance of the bilinear form, in contrast to our situation.} These equations of motion are known to have a Lax representation, and we can readily adapt it to our case.

The standard Lax ansatz in the relativistic case takes the form\footnote{See e.g. \cite{Arutyunov:2009ga}, but note that our $A$ has opposite sign.}
\begin{equation*}
\mathcal{L}_\alpha = \ell_0 A_\alpha^{(0)} + \ell_1 A_\alpha^{(1)} + \ell_2 \gamma_{\alpha\beta}\varepsilon^{\beta \gamma} A_\gamma^{(1)},
\end{equation*}
where the curvature of $\mathcal{L}$, $\mathcal{F}(\mathcal{L})$, can be split over the grading of the algebra. Grade zero flatness of $\mathcal{L}$ requires
\begin{equation}
\mathcal{F}(\mathcal{L})^{(0)} = \varepsilon^{\alpha\beta}(2\ell_0\partial_\alpha A_\beta^{(0)} + \ell_0^2[A_\alpha^{(0)},A_\beta^{(0)}]+(\ell_1^2-\ell_2^2)[A_\alpha^{(1)},A_\beta^{(1)}] = 0,
\end{equation}
which follows since $A$ itself is flat, provided we take $\ell_0 = 1$, and $\ell_1^2 - \ell_2^2 = 1$, leaving a single free (spectral) parameter in the coefficients $\ell_1$ and $\ell_2$. In grade one we then find
\begin{equation}
\mathcal{F}(\mathcal{L})^{(1)} = \ell_1 (\mathcal{F}(A))^{(1)} + \ell_2(\mathcal{E})\,.
\end{equation}
We see that since $A$ is flat, and $\ell_2$ a nontrivial function of the spectral parameter, flatness of the Lax connection is equivalent to the equations of motion.

In our case, $J$ differs by $A$ only in grade one, and by a term that commutes with everything in grade one. In other words, the modification does not affect grade zero. Any modification of this type can be accounted for in the Lax ansatz simply by replacing $A$ by $J$ in the $\ell_2$ term, i.e. by taking
\begin{equation}
\label{eq:lax}
\mathcal{L}_\alpha = \ell_0 A_\alpha^{(0)} + \ell_1 A_\alpha^{(1)} + \ell_2 \gamma_{\alpha\beta}\varepsilon^{\beta \gamma} J_\gamma^{(1)}.
\end{equation}
If we consider flatness of this Lax connection, in grade zero we find the same as before -- i.e. the same solution for the $\ell_i$ admitting a spectral parameter -- since our modification does not affect grade zero, and $A$ is still flat. In grade one, we do not modify the $\ell_1$ term, which hence still vanishes due to flatness of $A$, while the $\ell_2$ term precisely picks up the desired replacement of $A$ by $J$ to get the equations of motion \eqref{eq:eom_compact}. Of course, unlike the relativistic case, our equations of motion still come supplemented with the constraints (\ref{l+_eom},\ref{l-_eom}).

\section{The NR AdS string}\label{sec:NR_AdS}

The NR AdS string, as obtained in \cite{Gomis:2005pg} from the AdS$_5\times$S$^5$ string, has coordinate Polyakov action\footnote{This and other choices of coordinates are discussed in detail in \cite{Fontanella:2021hcb}.}
\begin{equation}\label{eq:NRAdS}
\begin{aligned}
S &= - \frac{T}{2} \int \dd^2 \sigma \, \bigg[\gamma^{\alpha\beta}\bigg( x^a x_a (-\partial_{\alpha} x^0 \partial_{\beta} x^0 + \cos^2 x^0 \partial_{\alpha} x^1 \partial_{\beta} x^1)
+\partial_{\alpha} x^a \partial_{\beta} x_a + \partial_{\alpha} x^{a'} \partial_{\beta} x_{a'} \bigg)\\
&+ \varepsilon^{\alpha\beta} \bigg( - (\lp e_{\alpha}{}^+ + \lm e_{\alpha}{}^-) \partial_{\beta} x^0 + (\lp e_{\alpha}{}^+ - \lm e_{\alpha}{}^-) \cos x^0 \partial_{\beta} x^1 \bigg) \bigg] \ ,
\end{aligned}
\end{equation}
where $x^0, x^1$ are longitudinal coordinates originating from AdS$_5$, while $x^a$ and $x^{a'}$, with $a, b, ... = 2, 3, 4$ and $a', b', ... = 1, ..., 5$, are transverse coordinates originating from AdS$_5$ and S$^5$ respectively, which are contracted with $\delta_{ab}$.
The corresponding NG action in static gauge is given by
\begin{equation}\label{eq:NRAdSNG}
 S_{NG} =  - \frac{T}{2} \int \dd^2 \sigma \, \sqrt{-\tau}\bigg( \tau^{\alpha\beta} \partial_{\alpha} x^a \partial_{\beta} x_a + \tau^{\alpha\beta} \partial_{\alpha} x^{a'} \partial_{\beta} x_{a'}  + 2 x^a x_a \bigg) \ ,
\end{equation}
where $\tau_{\alpha\beta} =$ diag$(-1, \cos^2 \tau)$ is the AdS$_2$ metric. The above action describes eight free scalar fields in AdS$_2$, five massless and three with mass two. This NR string can be described as a coset model similarly to the flat space case. Here we would like to briefly summarise the relevant modifications.

\subsection{Symmetry algebra}

To start we need to determine our symmetry algebra. Unlike the NR flat space string, the global symmetries of the NR AdS string have not been explicitly determined. It is however readily possible to determine a sufficient finite dimensional (sub)algebra for our construction. One way to do so is to consider the string Bargmann algebra that we used for the NR flat space string, as arising via Lie algebra expansion of the Poincar\'e symmetry of the relativistic flat space string, see e.g. \cite{Harmark:2019upf}. Applying the same technique to the $\mathfrak{so}(4,2) \oplus \mathfrak{so}(6)$ of the relativistic AdS$_5\times$S$^5$ string, we obtain a new algebra given by the direct sum of an extended string Newton-Hooke algebra, and an Euclidean algebra, see Appendix \ref{app:Liealgebraexp} for details.

The generators of the extended string Newton-Hooke algebra are the same generators of the string Bargmann algebra, where now $A, B, \ldots =0,1$ and $a, b, \ldots  = 2,3,4$. They satisfy the commutation relations (\ref{eq:Bargmann}), plus
\begin{equation}
\label{string_NH}
	[H_A,Z_B]  = -\varepsilon_{AB} Z \,,\qquad
	[H_A, H_B] = -\varepsilon_{AB} M \, ,\qquad
	[H_A, P_b] = G_{Ab} \ .
\end{equation}
The generators of the five-dimensional Euclidean algebra are the spatial translations $P_{a'}$ and the spatial rotations $J_{a'b'}$, where $a', b',\ldots  = 1, \ldots , 5$, with non-trivial commutation relations
\begin{equation}
\label{Euclidean}
[P_{a'}, J_{b'c'}] = \delta_{a'b'} P_{c'} - \delta_{a'c'} P_{b'} \, , \qquad
[J_{a'b'}, J_{c'd'}] = \delta_{b'c'} J_{a'd'} - \delta_{a'c'} J_{b'd'} + \delta_{a'd'} J_{b'c'} - \delta_{b'd'} J_{a'c'} \, .
\end{equation}

Our coset description is now based on
  \begin{subequations}
 	\begin{eqnarray}
 	\mathfrak{g} &=& \text{span} \{H_A, P_a, M, J_{ab}, G_{Aa}, Z_A, Z, P_{a'}, J_{a'b'} \} \ , \\
 	\mathfrak{h} &=& \text{span} \{M, J_{ab}, G_{Aa}, Z_A, Z, P_{a'}, J_{a'b'} \} \ .
 	\end{eqnarray}
 \end{subequations}
where we can give $\mathfrak{g}$ a $\mathbb{Z}_2$ grading corresponding to
\begin{equation}
\mathfrak{g}^{(0)} = \text{span} \{ M, J_{ab}, G_{Aa}, Z, J_{a'b'} \}\ , \qquad
\mathfrak{g}^{(1)} = \text{span} \{ H_A, P_a, Z_A, P_{a'} \}  \, ,
\end{equation}

\subsection{Bilinear form and action}

Our $\mathfrak{g}^{(0)}$-invariant bilinear form is now
\begin{subequations}\label{inner_product_AdS}
	\begin{align}
	\langle P_a, P_b \rangle &=  \omega_1  \delta_{ab} \ , &
	\langle H_{\pm}, Z_{\mp} \rangle &=  -\omega_1/2 \ , \\
	\langle H_+, H_- \rangle &= \omega_2 \ , &
	\langle M, M \rangle &= \omega_3 \ , \\
	\langle J_{ab}, J_{cd} \rangle &= \omega_4 \delta_{[a[c} \delta_{b]d]}  \ , &
	\langle J_{a'b'}, J_{c'd'} \rangle &= \omega_5 \delta_{[a'[c'} \delta_{b']d']}  \ , \\
	\langle P_{a'}, P_{b'} \rangle &=  \omega_6  \delta_{a'b'} \ ,
	\end{align}
\end{subequations}
which has the same bonus adjoint invariance (\ref{bonus}) as before, and again is grade compatible. The NR AdS string action is still given by equation (\ref{eq:NR_coset_action}), provided that $\omega_6 = \omega_1 = 1$. Gauge invariance of this action follows as before, with the same transformation rules for the Lagrange multipliers.\footnote{The new terms generated from the transformation of the MC form by the new nonzero commutators \eqref{string_NH} do not affect the Lagrangian due to the structure of the bilinear form.}

With the coset representative
\begin{equation}
\label{eq:NRAdScosetrep}
g = g_{l} g_{t}, \qquad g_l = e^{x^1 H_1} e^{-x^0 H_0}, \qquad  g_t = e^{x^a P_a + x^{b'} P_{b'}} \ ,
\end{equation}
we find the MC form
\begin{equation}
A =  g_t^{-1} A_l g_t + A_t \ ,
\end{equation}
where
\begin{equation}
A_l = g_l^{-1} \dd g_l = e^A H_A + \omega M = - \dd x^0 H_0 + \cos{x^0} \dd x^1 H_1 + \sin{x^0} \dd x^1 M\ ,
\end{equation}
is the longitudinal MC form corresponding to a coset representative for AdS$_2$, i.e. $e$ is the AdS$_2$ vielbein and $\omega$ the spin connection. The transverse MC form $A_t$ is
\begin{equation}
A_t = \dd x^a P_a + \dd x^{b'} P_{b'}\,,
\end{equation}
corresponding to a vielbein for $\mathbb{R}^8$. Using the explicit commutation relations, we readily find
\begin{equation}
\label{eq:MCforNRAdS}
A = A_t + A_l + e^A (x^a G_{Aa} +  \tfrac{1}{2} x^a x_a Z_A).
\end{equation}
Substituting this in the action we find the coordinate action \eqref{eq:NRAdS} above, the extra $x^a x_a$ terms arising precisely from $A_t$ combining with the extra $Z_A$ contribution in \eqref{eq:MCforNRAdS}.\footnote{Here we used the coordinates of \cite{Gomis:2005pg}. Alternatively, e.g. the coset representative $g_l = e^{t H_0} e^{x H_1}$ gives AdS$_2$ in global coordinates.}

\subsection{Equations of motion and Lax representation}

Coming next to the equations of motion, the extended string Newton-Hooke and the string Bargmann algebras differ by the commutation relations (\ref{string_NH}). These relations, however, can only contribute to MC components associated with the generators $M$, $Z$, and $G_{Ab}$. As these components do not appear in the action, there is no extra contribution to the equations of motion, which are again given by eqs. (\ref{Coset_eom}) or (\ref{eq:eom_compact}).\footnote{Curiously, (\ref{eq:eom_compact}) now also incorporates \eqref{M_eom}, albeit in its $Z$ component. This feature is not crucial however, since this equation of motion is proportional to those of the Lagrange multipliers.} Since gauge symmetry is formally the same as in the flat space case,\footnote{In terms of the full MC form there is a difference in the transformation of the $Z$ component, however this does not affect the action.} the Noether identities (\ref{Noether_Z_A}), (\ref{Noether_M}), again appropriately reduce the number of independent equations of motion.

Finally, coming to the Lax representation of the equations of motion, we can again take the ansatz \eqref{eq:lax}. However, now the difference between $J$ and $A$ no longer commutes with everything in grade one, since $[Z_A,H_B]\neq0$. Fortunately, the extra term, $\varepsilon^{\alpha\beta} [A^{(1)}_\alpha,\Lambda_\beta]$, only picks up a contribution from $A^{H_\pm}$ which is proportional to the equations of motion  \eqref{M_eom}, itself proportional to those of the Lagrange multipliers. Hence on solution of these constraints, our Lax representation still encodes the equations of motion (\ref{eq:eom_compact}).

It is interesting to note that the Lax connection \eqref{eq:lax} in the coordinates of eqs. (\ref{eq:NRAdScosetrep}-\ref{eq:MCforNRAdS}), upon solving the constraints in terms of the worldsheet zweibein and imposing static gauge, still depends on the Lagrange multiplier fields in a nontrivial way. As a result, this Lax connection encodes the equations of motion of the NG action \eqref{eq:NRAdSNG}, but only when taking into account the equations of motion for the longitudinal fields $x^0$ and $x^1$ of \eqref{eq:NRAdS}, that are gauge fixed to arrive at \eqref{eq:NRAdSNG}.\footnote{We could formally solve these equations for the Lagrange multiplier fields, but they nontrivially depend on the massive transverse field configuration. For the flat space NR string the situation is similar, but we can drop the Lagrange multiplier terms since $\lambda_\pm=0$ is a solution of the corresponding equations of motion. Dropping further inconsequential constant terms, this gives the obvious Lax connection $\mathcal{L} = \ell_1 \dd x^a P_a - \ell_2\star \dd x^a P_a$, with $\ell_1^2 - \ell_2^2 = 1$, for free massless fields in $\mathbb{R}^{1,1}$.}

\section{Conclusions}

In this paper we have given a coset formulation for the nonrelativistic flat space and AdS strings of \cite{Gomis:2000bd} and \cite{Gomis:2005pg} respectively, manifesting a finite dimensional part of the symmetries of these models. Using this formulation we were moreover able to give a Lax representation of the associated equations of motion.
While NR strings are typically referred to as a limit of relativistic strings, the actual procedure to arrive at them is not a straightforward limit, and leads to the introduction of new Lagrange multiplier fields, see appendix \ref{app:SNCreview}. For this reason, our coset formulation and in particular Lax representation do not simply follow as a limit of the corresponding relativistic ones.

Our formulation is intended to enable a systematic investigation of integrability of the NR flat space and AdS strings. There are various natural questions to pursue here. For one, here we focussed purely on the bosonic sector of the models, and it is important to extend our formulation to incorporate fermions. Next, while we have found a Lax representation of the equations of motion, this direction certainly deserves more study, in particular in a Hamiltonian framework. Moreover, with a coset formulation of these models, we may be able to define integrable deformations of these models, similar to those for the relativistic AdS$_5\times$S$^5$ string \cite{Delduc:2013qra}, see e.g. \cite{Hoare:2021dix} for an excellent pedagogical review. Since such deformations are typically built on the symmetries of a model, it would be interesting to see if the infinite dimensional symmetries of the NR flat space (and presumably AdS) string, that are not manifested in our coset approach, can still be involved here. Next, given that NR strings are known to be T dual to relativistic strings \cite{Bergshoeff:2018yvt}, and as a canonical transformation T duality preserves integrability, it could be enlightening to contrast these two descriptions from an integrability point of view.

It would also be interesting to go beyond the two models we considered here, and study coset model descriptions and integrability for NR version of other strings. Here the AdS$_3\times$S$^3\times$S$^1$ superstring, see e.g. \cite{Sfondrini:2014via}, immediately comes to mind, given its $\mathfrak{d}(2|1;\alpha)$ symmetry which may yield some new structure in the NR limit. Another interesting case would be the AdS$_3$ string supported by mixed NSNS and RR flux \cite{Cagnazzo:2012se}, assuming the B field can be nontrivially taken along in the NR limit.

Next, there is an interesting connection between NR strings and relativistic strings expanded near minimal surfaces associated with Wilson loops on the boundary \cite{Giombi:2017cqn}, where the NR string action appears at leading order \cite{Gomis:2005pg}.\footnote{There is evidence that this connection can be generalised to p-branes \cite{Sakaguchi:2006pg,Drukker:2020swu}.} It would be interesting to explore whether higher order terms may be incorporated in a would-be NR string action and whether the Lax representation can be correspondingly adapted.

Beyond integrability, it is important to understand how our formulation can be incorporated in the general TSNC framework for NR strings.

\section*{Acknowledgments}

We would like to thank Eric Bergshoeff, Jelle Hartong, Marc Magro, Niels Obers, and Arkady Tseytlin for helpful discussions and Gerben Oling and Arkady Tseytlin for comments on the draft. The work of AF was, and the work of ST continues to be supported by the German Research Foundation (DFG) via the Emmy Noether program ``Exact Results in Extended Holography''. Part of AF's work was completed while at the ITMP. AF thanks Lia for her permanent support. ST is supported by LT.


\setcounter{section}{0}
\setcounter{subsection}{0}
\setcounter{equation}{0}

\begin{appendices}

\section{Conventions and Identities}
\label{app:Conventions}

Here we give our conventions, and collect some identities used to show gauge invariance of the action, and derive Noether identities.

For a generic object $\mathcal{O}^A$, we define its light-cone combinations as
\begin{equation}\label{LC_comb}
\mathcal{O}^{\pm} \equiv \mathcal{O}^0 \pm \mathcal{O}^1 \ , \qquad\qquad
\mathcal{O}_{\pm} \equiv \frac{1}{2}\left( \mathcal{O}_0 \pm \mathcal{O}_1 \right) \ .
\end{equation}
The longitudinal Minkowski metric then has non-vanishing components $\eta_{+-} = -1/2$ and $\eta^{+-} = -2$. We take $\varepsilon^{01} = - \varepsilon_{01} = + 1$ for $\varepsilon^{\alpha\beta}$, $\varepsilon^{\sf ab}$ and $\varepsilon^{AB}$. In light-cone components $\varepsilon_{+-} = \frac{1}{2}$, $\varepsilon^{+-} = -2$.

The Hodge dual of a $p$-form $\omega = \frac{1}{p!} \omega_{\mu_1 \cdots \mu_p} \dd x^{\mu_1}\wedge \cdots \wedge \dd x^{\mu_p}$ is
\begin{equation}
\star \omega = \frac{\sqrt{|g|}}{p!(D-p)!} \omega_{\mu_1 \cdots \mu_p} \varepsilon^{\mu_1 \cdots \mu_p}{}_{\nu_{p+1} \cdots \nu_D}  \dd x^{\nu_{p+1}}\wedge \cdots \wedge \dd x^{\nu_D} \ .
\end{equation}
We denote symmetrization of indices by round brackets $(...)$ and anti-symmetrization by square ones $[...]$, including a $1/p!$ for $p$ indices. 	

Coming to identities, taking the worldsheet zweibein $e_{\alpha}{}^{\sf a}$ in light-cone components
\begin{equation}
e_{\alpha}{}^{\pm} = e_{\alpha}{}^0 \pm e_{\alpha}{}^1 \ , \qquad\qquad
e^{\alpha}{}_{\pm} = \frac{1}{2} \left( e^{\alpha}{}_0 \pm e^{\alpha}{}_1 \right) \ ,
\end{equation}
the worldsheet metric becomes
\begin{equation}
h_{\alpha\beta} = - e_{(\alpha}{}^+ e_{\beta)}{}^- \ , \qquad\qquad
h^{\alpha\beta} = - 4 e^{(\alpha}{}_+ e^{\beta)}{}_- \ .
\end{equation}
The determinant of the zweibein is
\begin{equation}
e = \sqrt{-h} = \text{det}(e_{\alpha}{}^{\sf a}) = e_{\tau}{}^0 e_{\sigma}{}^1 - e_{\tau}{}^1 e_{\sigma}{}^0 \ .
\end{equation}
The inverse zweibein $e^{\alpha}{}_{\sf a}$ is given in terms of $e_{\alpha}{}^{\sf a}$ via
\begin{equation}
e^{\alpha}{}_{\sf a} = - \frac{1}{e} \varepsilon^{\alpha\beta}\varepsilon_{\sf ab} e_{\beta}{}^{\sf b} \ ,
\end{equation}
In particular, the above equation implies
\begin{equation}
e^{\alpha}{}_+ = - \frac{1}{2 e} \varepsilon^{\alpha\beta} e_{\beta}{}^- \ , \qquad\qquad
e^{\alpha}{}_- = \frac{1}{2 e} \varepsilon^{\alpha\beta} e_{\beta}{}^+ \ .
\end{equation}
Products of zweibeine satisfy the identities
   \begin{subequations}
 	\begin{align}
 	e^{\alpha}{}_+ e^{\beta}{}_- &= - \frac{1}{4} h^{\alpha\beta} - \frac{1}{4 e} \varepsilon^{\alpha\beta} \ , &
 	e^{\alpha}{}_- e^{\beta}{}_+ &= - \frac{1}{4} h^{\alpha\beta} + \frac{1}{4 e} \varepsilon^{\alpha\beta} \ , \\
 	e_{\alpha}{}^+ e_{\beta}{}^- &= - h_{\alpha\beta} + e \,\varepsilon_{\alpha\beta}\ ,  &
 	e_{\alpha}{}^- e_{\beta}{}^+ &= - h_{\alpha\beta} - e \,\varepsilon_{\alpha\beta} \ , \\
 	e^{\alpha}{}_+ e_{\alpha}{}^+ &= e^{\alpha}{}_- e_{\alpha}{}^- = 1 \ , &
 	e^{\alpha}{}_+ e_{\alpha}{}^- &= e^{\alpha}{}_- e_{\alpha}{}^+ = 0 \ .
 	\end{align}
\end{subequations}
For the SNC gauge parameters $\sigma^{AB}$ we have
\begin{equation}
\sigma_{[10]} = \frac{1}{2} \sigma_{+-} = - \frac{1}{2} \sigma_{-+} \ , \quad
\sigma_{(01)} = \frac{1}{4}(\sigma_{++} - \sigma_{--}) \ , \quad
\sigma_{00} = \sigma_{11} = \frac{1}{4}(\sigma_{++} + \sigma_{--}) \ .
\end{equation}

\section{Review of (torsionless) SNC strings}
\label{app:SNCreview}

The Polyakov action for a string propagating on a (torsionless) SNC background is \cite{Bergshoeff:2018yvt}
\begin{equation}
\label{NR_Polyakov_action}
S^{\text{Poly}} = - \frac{T}{2} \int \dd^2 \sigma \, \bigg( \gamma^{\alpha\beta}\partial_{\alpha} X^{\mu} \partial_{\beta} X^{\nu} H_{\mu\nu} + \varepsilon^{\alpha\beta} (\lp e_{\alpha}{}^+ \tau_{\mu}{}^+ + \lm e_{\alpha}{}^- \tau_{\mu}{}^- )\partial_{\beta}X^{\mu}  \bigg) \ .
\end{equation}
where the notation is explained below (\ref{eq:NRflat}).
The couplings entering in this action are determined by non-invertible vielbeine  $\{\tau_{\mu}{}^A, m_{\mu}{}^A, E_{\mu}{}^a\}$ of a String Newton-Cartan manifold. The tangent space indices are split into longitudinal $A, B, \ldots = 0, 1$ and transverse $a, b, \ldots = 2, \ldots, D-1$. Longitudinal indices are raised and lowered with $\eta_{AB}$, transverse ones with $\delta_{ab}$.

The SNC vielbeine $\tau_{\mu}{}^A$ and $E_{\mu}{}^a$, although not invertible, satisfy projected invertibility conditions,
\begin{subequations}
	\begin{align}
	\label{projected_invert_1}
	\tau_{\mu}{}^A \tau^{\mu}{}_B &= \delta^A_B \ , &
	\tau_{\mu}{}^A \tau^{\nu}{}_A + E_{\mu}{}^a E^{\nu}{}_a = \delta^{\nu}_{\mu} \ , \\
	\label{projected_invert_2}
	E_{\mu}{}^a E^{\mu}{}_b &= \delta^a_b \ , &
	\tau^{\mu}{}_A E_{\mu}{}^a = E^{\mu}{}_a \tau_{\mu}{}^A = 0 \ .
	\end{align}
\end{subequations}
There are two useful quantities we can construct out of the SNC vielbeine. One is the \emph{longitudinal metric} $\tau_{\mu\nu}$,
\begin{equation}
\tau_{\mu\nu} \equiv \tau_{\mu}{}^A \tau_{\nu}{}^B \eta_{AB} \ ,
\end{equation}
while the other one is the so-called \emph{boost invariant metric} $H_{\mu\nu}$,
 \begin{equation}
  H_{\mu\nu} \equiv  \big(\tau_{\mu}{}^A m_{\nu}{}^B + \tau_{\nu}{}^A m_{\mu}{}^B \big) \eta_{AB} + E_{\mu}{}^a E_{\nu}{}^b \delta_{ab}\ .
  \end{equation}

The equations of motion for the fields $\lambda_{\pm}$ are
\begin{equation}
\varepsilon^{\alpha\beta} e_{\alpha}{}^{\pm} \tau_{\mu}{}^{\pm} = 0\ ,
\end{equation}
which admit the solution
\begin{equation}
e_{\alpha}{}^{\pm} = \phi_{\pm}\tau_{\mu}{}^{\pm}\,
\end{equation}
where $\phi_{\pm}$ are arbitrary functions of the world-sheet coordinates. This implies that $h_{\alpha\beta} = \phi_+ \phi_- \tau_{\alpha\beta}$, where $\tau_{\alpha\beta} = \tau_{\mu\nu} \partial_{\alpha} X^{\mu} \partial_{\beta} X^{\nu}$, and leads to the NG form of the action,
\begin{equation}
S^{\text{NG}} = - \frac{T}{2} \int \dd^2 \sigma \, \sqrt{-\tau}\tau^{\alpha\beta}\partial_{\alpha} X^{\mu} \partial_{\beta} X^{\nu} H_{\mu\nu} \ .
\end{equation}

The action (\ref{NR_Polyakov_action}) is locally invariant under the SNC algebra, which is a non-central extension of the string Galilei algebra. Its generators are longitudinal boost $M$, longitudinal translations $H_A$, transverse rotations $J_{ab}$, transverse translations $P_a$, string-Galilei boosts $G_{Ab}$, and non-central extensions $Z_A$ and $Z_{AB}$, with the traceless condition $Z^A{}_A = 0$. The commutation relations are the same as in (\ref{eq:Bargmann}), except that the commutators $[G_{Aa}, G_{Bb}] = \delta_{ab} \varepsilon_{AB} Z$ and $[H_A, Z] = \varepsilon_A{}^B Z_B$ are replaced by
\begin{equation} \label{eq:SNC}
	\begin{aligned}
	{}[G_{Aa}, G_{Bb}] & = \delta_{ab} Z_{[AB]} \,, \\
	[H_A, Z_{BC}]  & = 2 \eta_{AC} Z_B - \eta_{BC} Z_A\,,\\
	[Z_{AB}, M]  & = \varepsilon_A{}^C Z_{CB} + \varepsilon_B{}^C Z_{AC}\,.\\
	\end{aligned}
\end{equation}
Infinitesimal transformations of SNC vielbeine (and associated spin connections) can be derived via the gauging procedure of the SNC algebra \cite{Bergshoeff:2019pij} . For this purpose we define the SNC algebra valued 1-form
\begin{equation}
\Theta_{\mu} \equiv  \tau_{\mu}{}^A H_A + E_{\mu}{}^a P_a + m_{\mu}{}^A Z_A + \Omega_{\mu} M + \frac{1}{2}\Omega_{\mu}^{ab} J_{ab} + \Omega_{\mu}^{Aa} G_{Aa} + n_{\mu}^{AB} Z_{AB} \ .
\end{equation}
and the algebra valued gauge parameter
\begin{equation}
\Pi = \Sigma M + \frac{1}{2}\Sigma^{ab} J_{ab} + \Sigma^{Aa} G_{Aa} + \sigma^A Z_A  + \sigma^{AB} Z_{AB} \ .
\end{equation}
We ignore the transformations associated with longitudinal and transverse translations, assuming they are traded for diffeomorphism via the standard procedure, see e.g. \cite{Hartong:2015zia}.
The gauge transformations of SNC vielbeine and spin connections are then given by
\begin{equation}
\delta \Theta_{\mu} = \partial_{\mu} \Pi + [\Theta_{\mu}, \Pi] \, ,
\end{equation}
which explicitly are
\begin{subequations} \label{infinitesimal_SNC}
\begin{eqnarray}
\delta \tau_{\mu}{}^A &=& -\Sigma \varepsilon^A{}_B \tau_{\mu}{}^B \ , \\
\delta E_{\mu}{}^a &=&  \Sigma_A{}^a \tau_{\mu}{}^A - \Sigma^{a}{}_{b} E_\mu{}^{b} \ , \\
	\delta m_\mu{}^A &=& \partial_{\mu} \sigma^A + \varepsilon^A{}_B \sigma^B \Omega_\mu - \Sigma \, \varepsilon^A{}_B m_\mu{}^B - \Sigma^{Aa} E_{\mu a} + \tau_\mu{}^B \sigma^A{}_B\ , \\
 \delta \Omega_\mu &=& \partial_{\mu} \Sigma\ , \\
 	\delta \Omega_\mu{}^{ab} &=& \partial_{\mu} \Sigma^{ab} - \Sigma^{ca} \Omega_\mu{}^{b}{}_{c} + \Sigma^{cb} \Omega_\mu{}^{a}{}_{c}\ , \\
 		\delta \Omega_\mu{}^{Aa} &=& \partial_{\mu} \Sigma^{Aa} - \Sigma \, \varepsilon^A{}_B \Omega_\mu{}^{Ba} + \varepsilon^A{}_B \Sigma^{Ba} \Omega_\mu - \Sigma^{a}{}_{b} \Omega_\mu{}^{Ab} + \Sigma^{A}{}_{b} \Omega_\mu{}^{ab} \ ,\\
 		\delta n_\mu{}^{AB} &=& \partial_{\mu} \sigma^{AB} - 2 \varepsilon_C{}^B \bigl( \Omega_\mu \, \sigma^{(AC)} - \Sigma \, n_\mu{}^{(AC)} \bigr) - 2 \Sigma^{[A}{}_{a} \Omega_\mu{}^{B]a}\,.
\end{eqnarray}
\end{subequations}
The action (\ref{NR_Polyakov_action}) is invariant under local transformations (\ref{infinitesimal_SNC}), with two further assumptions. For the $\sigma^A Z_A$ transformation we need to impose the zero torsion constraint, i.e. that the $H_A$ component of the curvature of $\Theta$ vanishes
\begin{equation}
\label{zero_torsion}
\left( \left. d \Theta + [\Theta,\Theta]\right) \right|_{H_A} =0,
\end{equation}
and we need to take the $\lambda_{\pm}$ fields to transform as
 \begin{subequations}\label{l_lb_gauge}
\begin{eqnarray}
\delta\lp &=& -\Sigma \lp + 2 e^{\alpha}{}_+  \left(\tilde{\mathcal{D}}_{\alpha}\sigma^- - \frac{1}{2}\tau_{\alpha}{}^+ \sigma_{++} \right)\ , \\
\delta \lm &=& \Sigma \lm - 2 e^{\alpha}{}_-  \left(\tilde{\mathcal{D}}_{\alpha}\sigma^+ - \frac{1}{2} \tau_{\alpha}{}^- \sigma_{--} \right) \ ,
\end{eqnarray}
\end{subequations}
where
\begin{equation}
\tilde{\mathcal{D}}_{\alpha} \sigma^A \equiv \partial_\alpha \sigma^A - \varepsilon_B{}^A \Omega_{\alpha} \sigma^B,
\end{equation}
under $\Sigma M$ and $\sigma^A Z_A$ transformations. The Lagrangian of eqn. (\ref{NR_Polyakov_action}) transforms under the infinitesimal action of $Z_A$ by a total derivative term, while its variation is exactly zero under the action of the remaining SNC generators.

\subsection*{Relation to the coset approach}

Although the local symmetry of the action (\ref{NR_Polyakov_action}) is the SNC algebra, to find a coset action and Lax representation it proves useful to restrict to the string Bargmann algebra, which is a subalgebra of the SNC algebra obtained by setting $Z_{++} = Z_{--}=0$, and identifying $Z_{+-} = \frac{1}{2} Z$. There is then a clear connection between the gauging procedure described above, and our coset construction.

The Maurer-Cartan form $A_{\mu}$ may be regarded as a particular case of the one form $\Theta_{\mu}$, and we can think of the various components of $A$ as corresponding to particular SNC fields. Since the MC form is flat by construction, however, we are automatically restricted to the subclass of SNC geometries with zero curvature of all types, including in particular the torsion constraint (\ref{zero_torsion}) represented by the $H_A$ components of the curvature.\footnote{Since complete flatness is automatically preserved by gauge transformations, we also do not need to modify the gauge transformation of $n_{\mu}$ to preserve the curvature constraints, as needed in general \cite{Bergshoeff:2019pij}.}

To make contact between the coset action (\ref{eq:NR_coset_action}) and the SNC string action (\ref{NR_Polyakov_action}), we can write (\ref{eq:NR_coset_action}) explicitly as
\begin{equation}
\label{NR_coset_flat_action}
S^{\mathcal{G}/\mathcal{H}} = \int \dd^2 \sigma \, \left( \mathcal{L}_{\text{kin}} + \mathcal{L}_{\text{WZ}} \right)\ ,
\end{equation}
where
 \begin{subequations}
 	\begin{align}
 \mathcal{L}_{\text{kin}} &= - \frac{T}{2} \gamma^{\alpha\beta} \left(
 \delta_{ab} A_{\alpha}^{P_a} A_{\beta}^{P_b}
 -  A_{\alpha}^{H_+} A_{\beta}^{Z_-}
 - A_{\alpha}^{H_-} A_{\beta}^{Z_+}   \right)  \ , \\
 \mathcal{L}_{\text{WZ}} &= - \frac{T}{2}  \varepsilon^{\alpha\beta} \left(
  \Lambda_{\alpha}^{Z_-} A_{\beta}^{H_+}
 + \Lambda_{\alpha}^{Z_+} A_{\beta}^{H_-}  \right) \ ,
	\end{align}
\end{subequations}
and identify the components of $A$ in terms of those of $\Theta$.

\subsection*{Connection with the relativistic string action}

The action (\ref{NR_Polyakov_action}) can also be derived by a limiting procedure from the relativistic string action,
\begin{equation}\label{rel_action}
S = - \frac{T}{2} \int \dd^2 \sigma \, \bigg( \gamma^{\alpha\beta}\partial_{\alpha} X^{\mu} \partial_{\beta} X^{\nu} G_{\mu\nu} + \varepsilon^{\alpha\beta}\partial_{\alpha} X^{\mu} \partial_{\beta} X^{\nu} B_{\mu\nu} \bigg) \ ,
\end{equation}
where $G_{\mu\nu}$ is the spacetime metric and $B_{\mu\nu}$ is the Kalb-Ramond 2-form field, which is assumed to be closed.\footnote{It is possible to generalise this procedure for a generic 2-form, and also including the dilaton \cite{Bergshoeff:2018yvt}.}


To consider a NR limit of this action we choose two spacetime coordinates to rescale, one of which is timelike.\footnote{The NR limit is not unique, and rescaling only the time-like coordinate can also provide meaningful string models \cite{Batlle:2016iel}.}. For curved backgrounds, this choice in general affects the resulting model, and not all choices lead to an action of the form (\ref{NR_Polyakov_action}). Symmetries of the relativistic background may indicate suitable rescalings, depending on the desired outcome.

Let us suppose that a suitable coordinate rescaling has been identified and that it induces the following rescaling of relativistic vielbein $\hat{E}_{\mu}{}^{\hat{A}}$ for the metric $G_{\mu\nu} = \hat{E}_{\mu}{}^{\hat{A}}\hat{E}_{\mu}{}^{\hat{B}} \hat{\eta}_{\hat{A}\hat{B}}$,
\begin{eqnarray}
\label{vielbeine_exp}
 \hat{E}_{\mu}{}^A = c \tau_{\mu}{}^A + \frac{1}{c} m_{\mu}{}^A + \mathcal{O}(c^{-2})\ , \qquad\qquad
 \hat{E}_{\mu}{}^a = e_{\mu}{}^a + \mathcal{O}(c^{-1}) \ ,
\end{eqnarray}
where $c$ is the scaling dimensionless parameter here assumed to be large. Plugging this into the relativistic action (\ref{rel_action}) gives
\begin{equation}
\label{S_NR_div}
S = - \frac{T}{2} \int \dd^2 \sigma \, \gamma^{\alpha\beta} \bigg( c^2 \partial_{\alpha} X^{\mu} \partial_{\beta} X^{\nu} \tau_{\mu\nu} + \partial_{\alpha} X^{\mu} \partial_{\beta} X^{\nu} H_{\mu\nu} + \varepsilon^{\alpha\beta}  \partial_{\alpha} X^{\mu} \partial_{\beta} X^{\nu} B_{\mu\nu}\bigg) + \mathcal{O}(c^{-2}) \ ,
\end{equation}
To compensate the divergent $c^2$ term, we can fine-tune the B-field as
 \begin{equation}
B_{\mu\nu} = c^2 \tau_{\mu}{}^A \tau_{\nu}{}^B \varepsilon_{AB} \ ,
\end{equation}
so that (\ref{S_NR_div}) becomes
\begin{equation}
\label{action_F2}
S  = - \frac{T}{2} \int \dd^2 \sigma \bigg( \, \gamma^{\alpha\beta}\partial_{\alpha} X^{\mu} \partial_{\beta} X^{\nu} H_{\mu\nu} + c^2 \, \gamma^{00} \mathcal{F}^A \mathcal{F}^B \tilde{\eta}_{AB} + \mathcal{O}(c^{-2}) \bigg)\ ,
\end{equation}
where
\begin{equation}
\label{F}
 \mathcal{F}^A = \tau_{\mu}{}^A \partial_0 X^{\mu} - \frac{1}{\gamma_{11}} \varepsilon^{AB} \tilde{\eta}_{BC} \tau_{\mu}{}^C \partial_1 X^{\mu} - \frac{\gamma_{01}}{\gamma_{11}} \tau_{\mu}{}^A \partial_1 X^{\mu} \ .
\end{equation}
The divergent term $c^2\mathcal{F}^A \mathcal{F}^B \tilde{\eta}_{AB}$ can be rewritten as a finite plus a subleading term, by introducing two Lagrange multiplier fields $\lambda_A$ as,
\begin{equation}
\label{rewriting}
- \frac{T c^2}{2} \int \dd^2 \sigma  \, \gamma^{00} \mathcal{F}^A \mathcal{F}^B \tilde{\eta}_{AB} = - \frac{T}{2} \int \dd^2 \sigma \bigg(  \lambda_A \mathcal{F}^A + \frac{1}{4 c^2 \gamma^{00}} \lambda_A \lambda^A\bigg) \ ,
\end{equation}
If we now take the strict limit $c\rightarrow \infty$, we find the NR action
\begin{equation}
\label{NR_action}
S^{NR} = - \frac{T}{2} \int \dd^2 \sigma \, \bigg( \gamma^{\alpha\beta}\partial_{\alpha} X^{\mu} \partial_{\beta} X^{\nu} H_{\mu\nu} + \lambda_A \mathcal{F}^A \bigg) \ .
\end{equation}
This is the form of the action given in \cite{Gomis:2005pg}, there for AdS strings. This is equivalent to the action (\ref{NR_Polyakov_action}) of \cite{Bergshoeff:2018yvt}, provided we choose our zweibein such that \cite{Fontanella:2021btt}
\begin{equation}
\label{positivity_Zweibeine}
e_1 \overline{e}_0 - e_0\overline{e}_1 \geq 0 \ .
\end{equation}

\section{Relation between Polyakov and MC e.o.m.}
\label{app:coordinateandMCeom}

Here we demonstrate the correspondence between the equations of motion obtained from the Polyakov form of the NR string action \ref{NR_Polyakov_action},\footnote{We work with the SNC form for convenience. It covers both our flat space and AdS case, and holds more generally.} and those derived from our coset space formulation.

First, the equations of motion for the $\lambda_{\pm}$ fields are clearly identical,
\begin{equation}
-\frac{2}{T} \frac{\delta S^{\text{Poly}}}{\delta \lambda_{\pm}} = \mathcal{E}^{\lambda_{\pm}} \ ,
\end{equation}
For the remaining equations, we first use the projected invertibility condition (\ref{projected_invert_1}) to split the equations of motion for $X^{\mu}$ derived from $S^{\text{Poly}}$ in two parts,
 \begin{equation}
\left\{ \tau^{\mu}{}_A \frac{\delta S^{\text{Poly}}}{\delta X^{\mu}} = 0,\  E^{\mu}{}_a \frac{\delta S^{\text{Poly}}}{\delta X^{\mu}} = 0 \right\} \qquad \Longleftrightarrow \qquad
\frac{\delta S^{\text{Poly}}}{\delta X^{\mu}} = 0 \ ,
\end{equation}
These parts can be related to the coset formulation equations of motion as
  \begin{subequations} \label{tau_E_projections}
	\begin{align}
\tau^{\mu}{}_A \frac{\delta S^{\text{Poly}}}{\delta X^{\mu}} &= - \frac{T}{2} \left( \mathcal{E}^{Z_A} + \tau^{\mu}{}_A m_{\mu}{}^B \mathcal{E}^{H_B} \right)\ , \\
E^{\mu}{}_a \frac{\delta S^{\text{Poly}}}{\delta X^{\mu}} &= - \frac{T}{2} \left( \mathcal{E}^{P_a} + E^{\mu}{}_a m_{\mu}{}^B \mathcal{E}^{H_B} \right) \ .
	\end{align}
\end{subequations}
Due to the Noether identities (\ref{Noether_Z_A}), $\mathcal{E}^{H_A}$ on the right hand side is not independent, but can be written purely in terms of $\mathcal{E}^{\lambda_{\pm}}$. This shows that the equations of motion derived from the NR Polyakov action match with the independent equations of motion derived of the coset formulation.

\section{Derivation of the NR AdS symmetry algebra}
\label{app:Liealgebraexp}

String Bargmann is a subalgebra of the (infinite dimensional) global symmetry algebra of NR strings in flat spacetime, but its commutation relations can also be derived via Lie algebra expansion from the Poincar\'e algebra \cite{Harmark:2019upf}. Here we apply the same method to the $\mathfrak{so}(4,2) \oplus \mathfrak{so}(6)$ symmetry of the relativistic AdS$_5\times$S$^5$ string, giving us an algebra that we will take to be (a finite dimensional subalgebra of) the symmetry algebra of our NR AdS string.

$\mathfrak{so}(4,2) \oplus \mathfrak{so}(6)$ is generated by the relativistic translations $P_{\hat{a}}$ and rotations $J_{\hat{a}\hat{b}}$ for the AdS$_5$ part, with $\hat{a}, \hat{b}, ... = 0, 1, ..., 4$, and spatial translations $P_{a'}$ and rotations $J_{a'b'}$ for the S$^5$ part, with $a', b', ... = 1, ..., 5$. As a first step we write down the commutation relations with all powers of $c$ reinstated,
\begin{subequations}\label{so(4,2)+so(6)}
	\begin{align}
	[P_{\hat{a}}, P_{\hat{b}}] &= \frac{1}{c^2} J_{\hat{a}\hat{b}} \ , &
	[P_{a'}, P_{b'}] &= - \frac{1}{c^2} J_{a'b'} \ , \\
	[P_{\hat{a}}, J_{\hat{b}\hat{c}}] &= 2 \eta_{\hat{a}[\hat{b}} P_{\hat{c}]} \ , &
	[P_{a'}, J_{b'c'}] &= 2 \delta_{a'[b'} P_{c']} \ , \\
	[J_{\hat{a}\hat{b}}, J_{\hat{c}\hat{d}}] &= 4\eta_{[\hat{b}[\hat{c}} J_{\hat{a}]\hat{d}]} \ , &
	[J_{a'b'}, J_{c'd'}] &= 4\delta_{[b'[c'} J_{a']d']} \ .
	\end{align}
\end{subequations}
We then decompose the index $\hat{a} = (A, a)$, with $A=0,1$ and $a=2, 3,4$, and identify the generators $J_{Ab} \equiv c \, G_{Ab}$ and $P_A \equiv \frac{1}{c} H_A$, while leaving the other generators unchanged. Next, by tensoring each generator $T_A$ with the polynomial ring in the variable $\zeta = \frac{1}{c^2}$, i.e. $T_A^{(n)} \equiv T_A \otimes \zeta^n$, where $n\ge 0$ is the level, the algebra (\ref{so(4,2)+so(6)}) gives us the graded algebra
\begin{subequations}\label{so(4,2)+so(6)_graded}
	\begin{align}
	[H_A^{(n)}, H_B^{(m)}] &= J_{AB}^{(n+m)} \ , &
	[H_A^{(n)}, P_b^{(m)}] &= G_{Ab}^{(n+m)}\ , \\
	[P_a^{(n)}, P_b^{(m)}] &= J_{ab}^{(n+m+1)} \ , &
	[H_A^{(n)}, J_{BC}^{(m)}] &= 2 \eta_{A[B} H_{C]}^{(n+m)} \ , \\
	[H_A^{(n)}, G_{Bc}^{(m)}] &= \eta_{AB} P_{c}^{(n+m)} \ , &
	[P_a^{(n)}, G_{Bc}^{(m)}] &= - \delta_{ac} H_B^{(n+m+1)} \ , \\
	[P_a^{(n)}, J_{bc}^{(m)}] &= 2 \delta_{a[b} P_{c]}^{(n+m)} \ , &
	[J_{AB}^{(n)}, G_{Cd}^{(m)}] &= 2 \eta_{C[B} G_{A]d}^{(n+m)}  \ , \\
	[G_{Ab}^{(n)}, G_{Cd}^{(m)}] &= - \eta_{CA} J_{bd}^{(n+m+1)} - \delta_{db} J_{AC}^{(n+m+1)} \ , &
	[J_{ab}^{(n)}, G_{Cd}^{(m)}] &= - 2 \delta_{d[b} G_{a]C}^{(n+m)}  \ , \\
	[J_{ab}^{(n)}, J_{cd}^{(m)}] &= 4\delta_{[b[c} J_{a]d]}^{(n+m)} &
	[P_{a'}^{(n)}, P_{b'}^{(m)}] &= - J_{a'b'}^{(n+m+1)}  \ , \\
	[P_{a'}^{(n)}, J_{b'c'}^{(m)}] &= 2 \delta_{a'[b'} P_{c']}^{(n+m)} \ , &
	[J_{a'b'}^{(n)}, J_{c'd'}^{(m)}] &= 4\delta_{[b'[c'} J_{a']d']}^{(n+m)}   \ .
	\end{align}
\end{subequations}
This infinite dimensional algebra can be truncated by setting all generators with $n\ge 2$ to zero, as they form an ideal. We can make a further truncation by keeping only the $n=0$ generators, together with $Z_{A} \equiv H_A^{(1)}$ and $Z \equiv J^{(1)}$, where we defined $J_{AB}^{(n)} \equiv - J^{(n)} \varepsilon_{AB}$. We also define $M \equiv J^{(0)}$. Dropping the superscripts, we find the algebra of eqs. (\ref{eq:Bargmann}), (\ref{string_NH}) and (\ref{Euclidean}).\footnote{There is a second way to derive this algebra via Lie algebra expansion, as discussed in the ``Newton-Hooke case'' of \cite{Fontanella:2020eje}. This would bring in a set of generators $\{ \accentset{(2)}{J}_{\underline{a}\underline{b}}, \accentset{(2)}{J}_{a'b'} \}$, which are absent in the above. Since these extra generators form an ideal, however, they can be truncated away.}

\section{Coset action based on the SNC algebra}
\label{app:SNCcosetconstruction}

As mentioned in section \ref{sec:symmetry_flat}, the global symmetry algebra of NR strings in flat spacetime is infinite dimensional, and there are (at least) two natural choices of finite dimensional subalgebra to attempt to base our coset construction on. Here we briefly discuss a coset construction based on the SNC algebra, highlighting the differences with the string Bargmann case discussed in the main text. Although there appears to be no obstruction to using the SNC algebra for the coset action, it is ill suited to find a Lax representation for the equations of motion.

For our coset model we now take $G$ to be the group generated by the SNC algebra of eqs. (\ref{eq:SNC}). This algebra is the same as the string Bargmann algebra, with the exception of $Z$ which is replaced by the traceless $Z_{AB}$. We again take $H$ to be generated by everything except the $H_A$ and $P_a$.

For the bilinear form, we now ask for invariance under the adjoint action of $\tilde{\mathfrak{h}} \equiv \mathfrak{h}\setminus \{Z_A, Z_{AB}\}$,\footnote{We now exclude also $Z_{AB}$ since the Lagrange multipliers transform nontrivially under it. Of course, strictly speaking we do not need to exclude the antisymmetric part of $Z_{AB}$.}, and find
\begin{subequations}\label{inner_product_SNC}
	\begin{align}
	\langle P_a, P_b \rangle &=  \omega_1  \delta_{ab} \ , &
	\langle H_{\pm}, Z_{\mp} \rangle &=  - \omega_1 / 2 \ , \\
	\langle H_+, H_- \rangle &= \omega_2 \ , &
	\langle M, M \rangle &= \omega_3 \ , \\
	\langle Z_{++}, Z_{--} \rangle &= \omega_4 \ , &
	\langle J_{ab}, J_{cd} \rangle &= \omega_5 \eta_{[a[c} \eta_{b]d]}  \ ,
	\end{align}
\end{subequations}
for arbitrary $\omega_i$. It still has the bonus invariance (\ref{bonus}), now for $x \in \{Z_{\pm}, Z_{AB} \}$. Gauge invariance of the action works as in the string Bargmann construction, now taking the Lagrange multipliers to transform as in (\ref{l_lb_gauge}) instead of as in section \ref{sec:NRflatspaceaction}, compensating the new MC gauge transformations.

The graded subspaces $\mathfrak{g}^{(0)}, \mathfrak{g}^{(1)}$ remain defined as in (\ref{eq:grading}), but with the replacement of $Z$ with $Z_{AB}$. In particular, now $\mathfrak{g}^{(0)}$ is no longer equal to $\tilde{\mathfrak{h}}$. Then, provided that $\omega_1 = 1$ and $\omega_2 = 0$, the action takes again the form (\ref{eq:NR_coset_action}).

The equations of motion are as in (\ref{Coset_eom}), except for (\ref{H_eom}) and (\ref{Hb_eom}), which now become
   \begin{subequations}\label{H_eom_SNC_both}
   	\begin{align}
   	& \xi^{H_+}: &\mathcal{E}^{Z_-} \equiv &\ \partial_{\alpha}(\gamma^{\alpha\beta} A_{\beta}^{Z_-} + \varepsilon^{\alpha\beta} \Lambda_{\beta}^{Z_-})  - \varepsilon^{\alpha\beta} A_{\alpha}^M \Lambda_{\beta}^{Z_-}  \notag \\
 \label{H_eom_SNC}
 & \hspace{3cm}& &+ \gamma^{\alpha\beta} \left( \delta_{ab} A_{\alpha}^{G_{-a}} A_{\beta}^{P_b} - A_{\alpha}^M A_{\beta}^{Z_-} - A_{\alpha}^{Z_{+-}} A_{\beta}^{H_-} - A_{\alpha}^{Z_{--}} A_{\beta}^{H_+}\right) = 0 \ , \\
 & \xi^{H_+}: &\mathcal{E}^{Z_+} \equiv &\ \partial_{\alpha}(\gamma^{\alpha\beta} A_{\beta}^{Z_+} + \varepsilon^{\alpha\beta} \Lambda_{\beta}^{Z_+})  + \varepsilon^{\alpha\beta} A_{\alpha}^M \Lambda_{\beta}^{Z_+}  \notag \\
  \label{Hb_eom_SNC}
& \hspace{3cm}& &+ \gamma^{\alpha\beta} \left( \delta_{ab} A_{\alpha}^{G_{+a}} A_{\beta}^{P_b}+ A_{\alpha}^M A_{\beta}^{Z_+} + A_{\alpha}^{Z_{+-}} A_{\beta}^{H_+} - A_{\alpha}^{Z_{++}} A_{\beta}^{H_-} \right) = 0 \ .
	\end{align}
\end{subequations}
Moreover, there are two additional equations of motion,
  \begin{subequations}\label{Z_AB_eom}
   	\begin{align}
 \label{Z++_eom}
 &\xi^{Z_{++}}:  &\mathcal{E}^{Z_{++}} \equiv &\ \gamma^{\alpha\beta} A_{\alpha}^{H_-} A_{\beta}^{H_-}  = 0 \ , \\
 \label{Z--_eom}
 &\xi^{Z_{--}}:  &\mathcal{E}^{Z_{--}} \equiv &\ \gamma^{\alpha\beta} A_{\alpha}^{H_+} A_{\beta}^{H_+}  = 0 \ .
	\end{align}
\end{subequations}
The Noether identities discussion applies as before, except that now $Z_{AB}$ gauge invariance, with traceless gauge parameters $\sigma^{AB}$, implies
\begin{subequations}
   	\begin{align}
\delta_{\sigma_{++}} S^{\mathcal{G}/\mathcal{H}} = \frac{\delta S^{\mathcal{G}/\mathcal{H}}}{\delta A_{\mu}^{Z_-}} \delta_{\sigma_{++}} A_{\mu}^{Z_-} + \frac{\delta S^{\mathcal{G}/\mathcal{H}}}{\delta \lambda_+} \delta_{\sigma_{++}} \lambda_+ &\overset{!}{=} 0 \ ,\\
\delta_{\sigma_{--}} S^{\mathcal{G}/\mathcal{H}} = \frac{\delta S^{\mathcal{G}/\mathcal{H}}}{\delta A_{\mu}^{Z_+}} \delta_{\sigma_{--}} A_{\mu}^{Z_+} + \frac{\delta S^{\mathcal{G}/\mathcal{H}}}{\delta \lambda_-} \delta_{\sigma_{--}} \lambda_- &\overset{!}{=} 0 \ ,
	\end{align}
\end{subequations}
which gives the non-trivial Noether identities
\begin{subequations} \label{Noether_Z_AB}
   	\begin{align}
   	\mathcal{E}^{Z_{--}} - 2 e^{\alpha}{}_+ A_{\alpha}^{H_+} \mathcal{E}^{\lambda_+} &= 0 \ , \\
   	\mathcal{E}^{Z_{++}} + 2 e^{\alpha}{}_- A_{\alpha}^{H_-} \mathcal{E}^{\lambda_-} &= 0 \ .
	\end{align}
\end{subequations}
These extra Noether identities allow us to eliminate the extra equations of motion (\ref{Z_AB_eom}), again reducing the equations of motion to the independent set (\ref{independent_set}).

In (\ref{H_eom_SNC_both}), the $A_{\alpha}^{Z_{\pm \pm}} A_{\beta}^{H_{\mp}}$ terms arise from commutators with opposite sign from all the other commutator terms. This unfortunately prevents us from rewriting the equations of motion in the compact form (\ref{eq:eom_compact}) and it makes it unclear whether the equations of motion have a Lax representation.

\end{appendices}


\bibliographystyle{nb}

\bibliography{NR_coset_arXiv_v3}

\end{document}